# Matter Under Extreme Conditions: The Early Years

R. Norris Keeler

6652 Hampton Park Court, McLean, VA 22101, USA
rnkeeler@verizon.net

Carl H. Gibson

University of California San Diego, La Jolla, CA 92093-0411, USA
cgibson@ucsd.edu, http://sdcc3.ucsd.edu/~ir118

**Abstract:** Extreme conditions in natural flows are examined, starting with a turbulent big bang. A hydro-gravitational-dynamics cosmology model is adopted. Planck-Kerr turbulence instability causes Planck-particle turbulent combustion. Inertial-vortex forces induce a non-turbulent kinetic energy cascade to Planck-Kolmogorov scales where vorticity is produced, overcoming $10^{113}$ Pa Planck-Fortov pressures. The spinning, expanding fireball has a slight deficit of Planck antiparticles. Space and mass-energy powered by gluon viscous stresses expand exponentially at speeds $>10^{25}$ $c$. Turbulent temperature and spin fluctuations fossilize at scales larger than $ct$, where $c$ is light speed and $t$ is time. Because "dark-energy" antigravity forces vanish when inflation ceases, and because turbulence produces entropy, the universe is closed and will collapse and rebound. Density and spin fossils of big bang turbulent mixing trigger structure formation in the plasma epoch. Fragmenting protosuperclustervoids and protoclustervoids produce weak turbulence until the plasma-gas transition give chains of protogalaxies with the morphology of turbulence. Chain galaxy clusters observed at large redshifts ~8.6 support this interpretation. Protogalaxies fragment into clumps, each with a trillion Earth-mass H-He gas planets. These make stars, supernovae, the first chemicals, the first oceans and the first life soon after the cosmological event.

## 1.     Introduction

Modern astrophysical concepts and cosmological models are linked to the properties of matter under extreme conditions of the early universe. When astrophysical phenomena began to be observed, (quasars, pulsars, burstars) one of the first to point out that data under extreme conditions were lacking was Ya. B. Zel'dovich.[1] Historically, Russian investigators had obtained pressures in the multi-megabar range for iron and other earth core elements.[2]  US investigators also worked on earth core geophysics, but at lower pressures. A key part of this work was the determination of the melting point of iron on the Hugoniot, determined by Brown and McQueen.[3]  This gives clues to the condition at the Earth inner-core outer-core interface.  The only significant work on Outer Core electrical conductivity was done by one of the authors (RNK)[4]. Only one definitive paper on hydrogen was produced, establishing its metallization at ~ 2.3 Mbar and ~ 300 Kelvin. This work was relevant to the large planets, and enabled the prediction of Saturn's magnetic field.[5]

All this work is hardly relevant to astrophysical conditions after the big bang.  In the 1960-70 time frame, Vladimir Fortov began to study the properties of dense, high temperature plasma.  It was long thought that these states could not be easily reached.  In doing this, Fortov created an entire new area of physics to be explored,[6] which now can be used in studying primordial conditions.

Observations of increasing redshift with distance support a big bang cosmology in the standard (concordance) cosmological model $\Lambda$CDMHC,[7] consistent with Einstein's general relativity theory.  However, Einstein's equations are known to fail at Planck length scales $L_P = (c^{-3}hG)^{1/2}$, where $c$ is light speed, $h$ is Planck's constant and $G$ is Newton's constant, as the Schwarzschild radius of general relativity approaches the Compton wavelength of quantum mechanics[8,9] where both theories break down.

Modern fluid mechanics,[10] includes the dynamics of turbulent combustion and fossil turbulence. Both concepts are required to adequately describe big bang mechanisms without resorting to a permanent cosmological constant $\Lambda$ "dark energy".  One prediction is that the plasma protogalaxies fragment into clumps of gas planets[11] that freeze to form the galaxy dark matter, as confirmed by Schild in quasar microlensing observations.[12]  New observations reveal a general failure of nearly all aspects of the standard model, including the idea that ~ 70% of the mass-energy of the universe is the "dark energy" implied by a permanent cosmological constant $\Lambda$.

Failures of the standard cosmological model result from oversimplifications of the basic conser-vation laws for momentum, mass, and energy. Fluids are assumed to be collisionless, inviscid and linear. The concept that the big bang results from a turbulent instability is thus dismissed at the outset. Modern turbulence[13] theory and fossil turbulence[14] theory are crucial to understanding the cosmological big bang and gravitational structure formation during the plasma and gas epochs up to the present time. Because natural fluids are invariably stably stratified, transport processes involve a complex interaction between turbulence and gravity. Turbulence cascades from small scales to large, causing internal wave motions that propagate through the fluid to induce secondary turbulence and mixing events. Persistent fossil turbulence motions and microstructures form as buoyancy forces convert turbulence energy, momentum, and angular momentum to non-turbulence.

## 2. Hydro-Gravitational-Dynamics Theory

Fluid mechanical influences on cosmology exist from the beginning. Einstein's equations require a cosmological constant,[15] to overcome the enormous gravitational forces at Planck length scales.[16] Conditions existing at the start of the big bang[17] are illustrated in Figure 1.

The big bang is attributed[18,19] to a paired Planck particle turbulence instability. The mechanism is similar to the pair production process of supernovae, where electrons bind with positrons to form positronium. If a Planck particle and Planck antiparticle form a spinning pair, then a prograde accretion of a Planck antiparticle results in the release of 42% of the rest mass energy of the accreted particle,[20] which is marginally bound at only 10% of the Schwarzschild radius. At Planck temperatures, all of this energy must be released by creation of new Planck particle pairs, which will have the spin of the mother pair. If this is repeated, a spinning, turbulent, Planck particle gas cylinder results with increasing Reynolds number as the mass-energy increases. Gravitational forces induce powerful Planck-Fortov pressures that must be overcome for the process to proceed. Two such antigravity (dark energy) mechanisms are shown in Fig. 1. Inertial vortex forces of secondary turbulence eddy streams are radially outward, as shown, and polar inertial forces of the irrotational Planck gas also induce negative stresses that extract mass-energy from the vacuum.[21]

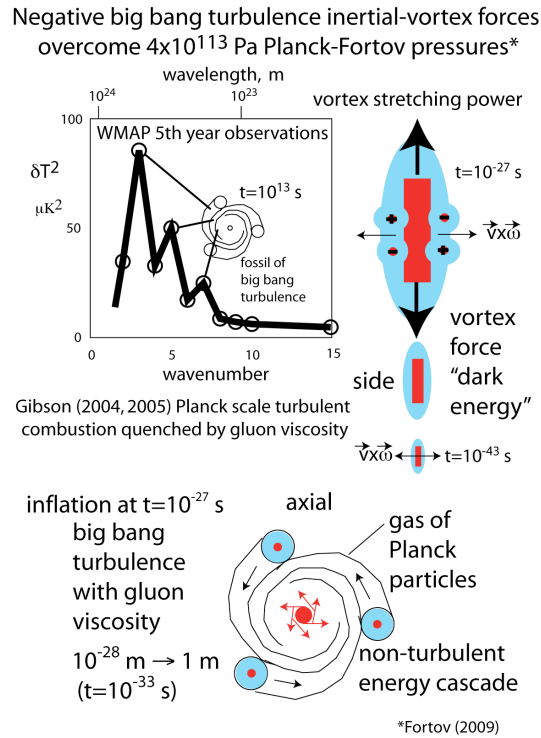

Fig. 1.  Negative big bang turbulence forces are needed to overcome extremely large Planck-Fortov pressures $L_P = c^7 h^{-1} G^{-2}$ of 4.6 $10^{113}$ Pa that resist the overturning inertial-vortex forces of turbulence eddies.

Secondary vortices wrap around the spinning turbulence fireball, and these cause the outward radial inertial-vortex forces that resist gravity in the big bang and produce "vortex force" mass-energy.  Evidence of this process can be seen in the cosmic microwave background temperature anisotropy spectrum and its largest scales shown in the insert from WMAP fifth year observations.[22]

The most powerful antigravitational force occurs when the expanding turbulent fireball cools to the strong force freeze-out temperature of ~ $10^{28}$ K, when quarks and gluons appear.  Because gluons transmit momentum over much larger scales than Planck particles, the expansion of the fireball is strongly resisted by gluon-viscosity.  Thus an enormous amount of work is done by the expansion against this negative stress that must be compensated by mass-energy production.  This is the mechanism of inflation.  In a time period of ~ $10^{-33}$ seconds, the fireball grows from ~ $10^{-28}$ meters to ~ 3 meter, giving a displacement speed of ~ $10^{25}$ $c$ and a universe of mass-energy $10^{98}$ kg.

Viscous forces prevent structure formation in the primordial plasma until the viscous gravitational length scale $L_{SC} = (\gamma\nu/\rho G)^{1/2}$ matches the scale of causal connection $ct$. This occurs at time $t = 10^{12}$ seconds, or 30,000 years, at a mass scale of $10^{46}$ kg, the mass of a supercluster.[23] The mechanism of gravitational structure formation at this time is one of fragmentation, since voids forming at density minima are favored by the expansion of space in the universe, whereas gravitational condensations are inhibited. The density of the fragmenting plasma $4 \times 10^{-17}$ kg m$^{-3}$ is preserved as a fossil of this event in the density of protogalaxies (PGs) and protoglobularstarclusters (PGCs). Completely empty voids are observed at scales exceeding $10^{25}$ m, versus only $10^{23}$ m of CDM halos empty of baryons expected from $\Lambda$CDMHC.[24]

A fossilized big bang turbulence vortex line is preserved as the axis of evil.[25] This is the direction of the dipole, quadripole, octopole etc. spherical harmonics of the CMB. This spin direction coincides with the spin vector of the Milky Way and other local galaxies. Spin partially explains the formation of weak turbulence by the expansion of the protosuperclustervoids, with Reynolds number $10^3$ smaller than that of the big bang.[26] Plasma turbulence fixes the morphology of protogalaxies that form at Kolmogorov-Nomura scales $\sim 10^{20}$ m before gas transition at $10^{12}$ s.

Because the morphology of the first galaxies matches that of the direct numerical simulation of Nomura-Post; that is, vortex tubes with spirals at the base, as shown in Figure 2, the protogalaxy scale $10^{20}$ meters is termed the Nomura scale $L_N$.

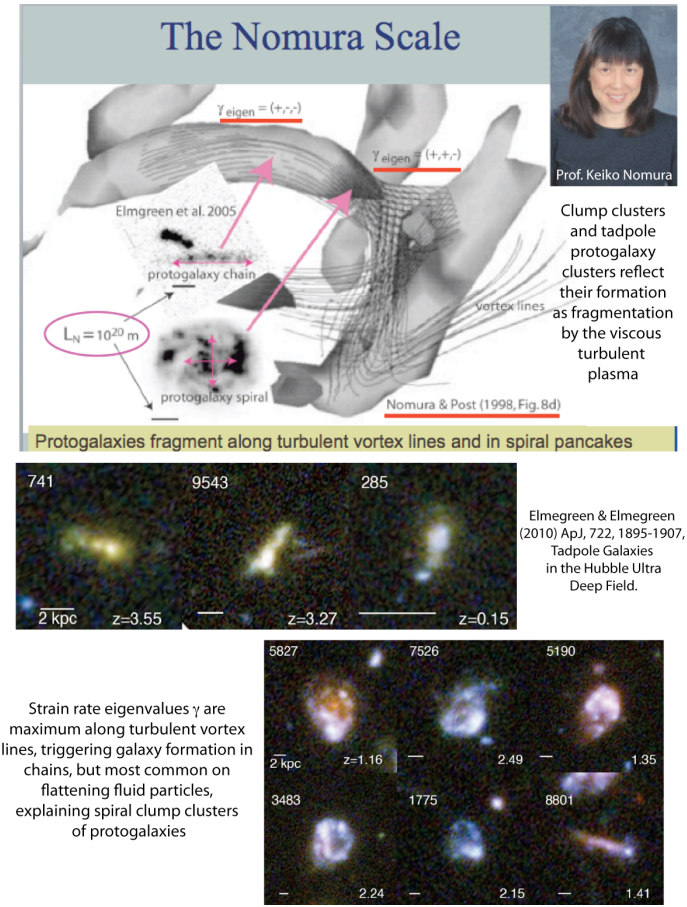

Fig. 2.  Nomura scales and turbulence morphology, reflected in the most distant clusters of protogalaxies formed in the last stages of the plasma epoch by weak plasma turbulence.

Protogalaxy fragmentation is expected along fossil plasma turbulence vortex lines where the magnitude of the rate of strain $\gamma$ is maximum and the eigenvalues are +,-,- (spaghetti).  This is observed in the Hubble Space Telescope Ultra Deep Field (HSTUDF) images as chains of bright clumps with scales $\sim L_N$.  Somewhat dimmer bright clumps with a spiral morphology are identified with the more common turbulence eigenvalues -,+,+ (pancakes).  Most of the extremely dim objects observed in such studies as the HSTUDF show the linear morphology of the tadpole objects identified by Elmegreen and Elmegreen.[27]

## 3.    Discussion

Evidence of chain and spiral galaxy clusters in Figure 2 is interpreted in terms of the big bang turbulence model of Figure 1.  According to big bang turbulence theory, there should be densely nested vortex lines during the initial instability period where Planck particles dominate and produce strong turbulence combustion before fossilization by gluon viscous forces that force inflation far beyond the horizon scale, fossilizing the turbulent temperature mixing structures. Inertial-vortex-forces of the turbulent big bang must overpower large gravitational Planck-Fortov pressures[28] for the eddies to overturn.  Fossil vorticity turbulence eddies and density fluctuations trigger gravitational structure formation during the plasma epoch to form protosuperclustervoids that expand at sonic speeds $c/3^{1/2}$.  The void sonic scale $ct/3^{1/2}$ explains the sonic peak observed in the CMB temperature anisotropies.

The plasma voids expand gravitationally as rarefaction waves that produce vorticity and weak turbulence by baroclinic torques.  Cluster and finally galaxy scale fragmentation occurs in the plasma epoch.  The non-baryonic dark matter is virtually irrelevant to the plasma structure formation because its diffusive scale is larger than $ct$.  The chain galaxy cluster morphology observed is that of the weak turbulence, as simulated by Nomura shown in Fig. 2.

## 4.      Conclusions

The standard model of cosmology conflicts with modern fluid mechanics[29,30,31] and with observations.[32,33,34,35]  Hydro-gravitational-dynamics (HGD) cosmology proposes a big bang turbulent combustion model capable of overcoming Planck-Fortov gravitational pressures[36] during a rapid cooling growth to the quark-gluon-viscosity temperature and the inflation epoch. A permanent dark energy constant $\Lambda$ is unnecessary and unlikely.  Supernova Ia evidence is explained as a systematic dimming error due to evaporation of ambient frozen primordial planets surrounding the supernova precursor.  Frictional turbulence dissipation suggests a closed universe.  A spin direction is observed at large scales, as expected from big bang turbulence but not from $\Lambda$CDMHC.  Voids form last in $\Lambda$CDMHC and are therefore much smaller than HGD voids that form early in the plasma epoch at near light speeds.  Observations of chains and spirals of protogalaxies at the Nomura scale support the turbulence morphology and viscous-gravitational scales expected from HGD cosmology of protogalaxies formed in the plasma epoch.

## Dedication

This paper is dedicated to Academician Vladimir E. Fortov on the occasion of the 50th anniversary of the founding of the Joint Institute for High Temperatures of the Russian Academy of Sciences.  Academician Fortov's work has provided a firm basis for astrophysical theories, and set bounds for new cosmological models.  It should also be noted that Academician Fortov's work has shown that many of the newer astrophysical phenomena observed can be explained by applying the results of his work, and do not require and new and exotic but unverifiable forces and phenomena.

Dr. R. Norris Keeler, Past President, International High Pressure Society

Professor Carl H, Gibson, Founding Editor in Chief, Journal of Cosmology

---

[1] Zeldovich, Ya, B, comments made at the COSPAR Conference, Budapest Hungary, June, 1980.

[2] Al'tshuler, L. V., Soviet Physics, Uspekikh **42** , 261(1999).

[3] Brown J. M., and R. G. McQueen, Journ. Geophys. Res. **91**, 7845(1986). The work of Ahrens on this problem has been shown to be wrong in Brown, J. M., "High Pressure Iron under Heated Debate, Deep Earth Dialog no. 7, SEDI News Letter Fall, 1993, p. 3.

[4] Keeler, R. N., and E. B. Royce, Proc. of the International School of Physic, Varenna, Italy, p 124 Course XLVIII, "Physics of High Energy Density", Academic Press, 1971. Royce was able to obtain conductivity of pure iron conductivity by using the skin effect associated with demagnetization effects, reference given herein.

[5] Hawke, R. S. et al., Phys Rev. Lett. 41, 994(1978).  Static investigators never achieved metallization, and the related work of Nellis has been discredited, in part by his own co-workers. Nellis' result was shown to not be metallization, but rather a high temperature-high density plasma phase transition, as first proposed by Teller, and later experimentally verified by Zel'dovich and Fortov.

[6] Keeler, R. N., "Some thoughts of an American Scientist on the dynamic high pressure work of academician Ya. B. Zel'dovich, Usp. Fiz. Nauk **165**, 595(1995).

[7] Peebles, P. J. E. and Ratra, Bharat 2003. The cosmological constant and dark energy, Rev. Mod. Phys., Vol. 75, No. 2.

[8] Gibson, C. H. 2004. The first turbulence and the first fossil turbulence, Flow, Turbulence and Combustion, 72, 161–179.

[9] Gibson, C. H. 2005. The first turbulent combustion, Combust. Sci. and Tech., 177:1049–1071, arXiv:astro-ph/0501416.

[10] Gibson, C. H. 1996. Turbulence in the ocean, atmosphere, galaxy and universe, Appl. Mech. Rev., 49, no. 5, 299–315.

[11] Ibid.

[12] Schild, R. 1996. Microlensing variability of the gravitationally lensed quasar Q0957+561 A,B, ApJ, 464, 125.

[13] Turbulence is defined as an eddy-like state of fluid motion where the inertial-vortex forces of the eddies are larger than any other forces that tend to damp the eddies out.  By this definition, turbulence always cascades from small scales to large since vorticity of an irrotational flow is produced at the Kolmogorov inertial-viscous scale; that is, at scales smaller than the Obukhov kinetic energy scale.

[14] Fossil turbulence is defined as a perturbation in any hydrophysical field produced by turbulence that persists after the fluid is no longer turbulent at the scale of the perturbation. Examples include jet airplane contrails and the axis of evil [6].

[15] Peacock, J. A. 2000. Cosmological Physics, Cambridge Univ. Press, UK., p 15.

[16] Fortov, V. E. 2009. Extreme states of matter on Earth and in space, Physics - Uspekhi 52 (6) 615 − 647.

[17] Gibson, C. H. (2010). Turbulence and turbulent mixing in natural fluids, Physica Scripta Topical Issue, Turbulent Mixing and Beyond (TMBW '09), (December issue), arXiv:1005.2772v4.

[18] Gibson, C. H. 2004, op. cit.

[19] Gibson, C. H. 2005., op. cit.

[20] Peacock, J. A. 2000.  Cosmological Physics, Cambridge Univ. Press, UK., p. 61

[21] Ibid., p. 26

[22] Gibson, C. H. (2010), op. cit.

[23] Gibson, C. H. 1996, op. cit.

[24] Gibson, C. H. (2010), op. cit.

[25] Schild, R. 1996, op. cit.

[26] Gibson, C. H. (2010), op. cit.

[27] Elmegreen, B. G. and Elmegreen, D. M. 2010.  Tadpole Galaxies in the Hubble Ultra Deep Field, Ap. J, 722, 1895-1907.

[28] Fortov, V. E. 2009. Extreme states of matter on Earth and in space, Physics - Uspekhi 52 (6) 615 − 647.

[29] Gibson, C. H. 2004, op. cit.

[30] Gibson, C. H. 2005, op. cit.

[31] Gibson, C. H. 1996, op. Cit.

[32] Schild, R. 1996, op. cit.

[33] Schild, R. 1996, Ibid.

[34] Gibson, C. H. (2010), op. cit.

[35] Elmegreen, B. G. and Elmegreen, D. M. 2010, op. cit.

[36] Fortov, V. E. 2009, op. cit.